\documentclass[conference]{IEEEtran} 
\usepackage{amsmath}
\usepackage{amsmath,amssymb,amsfonts}
\usepackage{cite}
\usepackage{mathtools}
\usepackage[binary-units=true]{siunitx}
\DeclareSIUnit{\belmilliwatt}{Bm}
\usepackage{amsfonts}
\usepackage{amssymb}
\usepackage{graphicx}
\usepackage{dsfont}
\usepackage{subfigure}
\usepackage{algorithm}
\usepackage{algpseudocode}
\usepackage{mathtools}

\usepackage{textcomp}
\usepackage[export]{adjustbox}
\usepackage{float}
\usepackage{booktabs}
\usepackage{multicol}
\usepackage{xparse}
\usepackage{color,soul}
\usepackage[bottom]{footmisc}
\usepackage[binary-units=true]{siunitx}
\DeclareSIUnit{\belmilliwatt}{Bm}
\DeclareSIUnit{\dBm}{\deci\belmilliwatt}
\DeclareSIUnit[per-mode=symbol,per-symbol=p]{\Bps}{\byte\per\second}

\sethlcolor{yellow}
\usepackage{algpseudocode}
\DeclareMathOperator{\E}{\mathbb{E}}

\usepackage{hyperref}

\makeatletter
\def\BState{\State\hskip-\ALG@thistlm}
\makeatother

\begin{document}

    \title{ AA-DRL: AoI-Aware Deep Reinforcement Learning Approach for D2D-Assisted Industrial IoT
}

\author{
\IEEEauthorblockN{Hossam Farag\IEEEauthorrefmark{1}, Mohamed Ragab\IEEEauthorrefmark{2}, and \v{C}edomir Stefanovi\'{c}\IEEEauthorrefmark{1}}
\IEEEauthorrefmark{1} Department of Electronic Systems, Aalborg University, Denmark\\
\IEEEauthorrefmark{2} Centre for Frontier AI Research, Agency for Science Technology and Research (A*STAR), Singapore\\
Email: hmf@es.aau.dk, mohamedr002@e.ntu.edu.sg, cs@es.aau.dk
}

	\maketitle
	
	\begin{abstract}
In real-time Industrial Internet of Things (IIoT), e.g., monitoring and control scenarios, the freshness of data is crucial to maintain the system functionality and stability. In this paper, we propose an AoI-Aware Deep Reinforcement Learning (AA-DRL) approach to minimize the Peak Age of Information (PAoI) in D2D-assisted IIoT networks. Particularly, we analyzed the success probability and the average PAoI via stochastic geometry, and formulate an optimization problem with the objective to find the optimal scheduling policy that minimizes PAoI. In order to solve the non-convex scheduling problem, we develop a Neural Network (NN) structure that exploits the Geographic Location Information (GLI) along with feedback stages to perform unsupervised learning over randomly deployed networks. Our motivation is based on the observation that in various transmission contexts, the wireless channel intensity is mainly influenced by distance-dependant path loss, which could be calculated using the GLI of each link. The performance of the AA-DRL method is evaluated via numerical results that demonstrate the effectiveness of our proposed method to improve the PAoI performance compared to a recent benchmark while maintains lower complexity against the conventional iterative optimization method. 
	\end{abstract}
\section{Introduction}\label{sec:intro}
The Internet of Things (IoT) technology evolves rapidly as a worldwide network of interconnected  intelligent devices that are capable of sensing, communicating, and processing to support a variety of applications, such as industrial monitoring, health monitoring and vehicular networks~\cite{iot}. Different from consumer IoT, Industrial IoT (IIoT) networks are characterized by strict communication requirements to maintain production efficiency and avoid safety-critical situation~\cite{Hossam}. IIoT networks are evolving from the typical star network configuration to Device-to-Device (D2D) communications where sensor-actuator pairs in propinquity communicate directly without evolving a central node (e.g., access point or a Base Station (BS)~\cite{D2D_IIoT}. In outband D2D communication~\cite{D2D_IIoT}, the D2D pairs operate in full frequency reuse model where the they communication over the same frequency band in uncoordinated fashion. For IIoT scenarios that comprise enormous D2D pairs, this incurs significant interference when D2D links are activated at the same time. Such uncoordinated access would cause noticeable degradation in the network performance in terms of delay, throughput and most importantly the information freshness.
Information freshness is crucial for a typical real-time control and monitoring scenario as it affects the derived intelligent and autonomous decisions and the system stability as well. For instance, in oil refineries, valve actuators should acquire timely monitoring of oil level to avoid oil tank spillage~\cite{oil_refinery}. Information freshness is quantified by Age-of-Information (AoI)~\cite{AoI}, a process-level metric which, from the receiver perspective, counts the time elapsed since the latest received information was generated. Efficient D2D links scheduling has been proved to effectively enhance various performance metrics~\cite{D2D_IIoT}. However, this approach is not necessarily effective to improve and optimize AoI. In addition, scheduling problems for  wireless networks with complex interference are usually non-convex and NP-hard~\cite{NP}. The conventional method for link scheduling first involves estimating the interference of channels and then optimizing the schedule using these estimates~\cite{complex}. Estimating Channel State Information (CSI) in densely deployed networks can be costly, and even achieving a near-optimal solution to the resulting optimization problem can be intricate. For instance, in network consisting of $N$ D2D links, $N^2$ CSI are required in the path-loss matrix within each coherence block, which corresponds to a computational complexity of at least $O(N^2)$. To that end, machine learning and artificial intelligence techniques are employed to find optimal schedules in dense networks~\cite{spatio}. 

In this work, we develop an AoI-Aware Deep Reinforcement Learning (AA-DRL) approach to minimize Peak AoI (PAoI) in dense D2D-assisted IIoT networks. The learning approach is based on the collected Geographic Location Information (GLI) to select the optimal scheduling policy. Our motivation is based on  the fact that in various transmission contexts, the wireless channel intensity is mainly influenced by distance-dependant path loss. Moreover, the pattern of interference within a network largely depends on how transmitters and receivers are positioned relative to one another. We derive the successful transmission probability and the average PAoI considering a preemptive queuing policy and spatial coupled interference between the D2D pairs. Then, we formulate the PAoI optimization problem under stationary randomized policy. We solve the non-convex scheduling problem via a proposed Neural Network (NN) structure where the scheduling policy is mapped to the GLI while explicit CSI is not required. The NN is trained via an unsupervised training process by utilizing the GLI along with feedback stages to obtain the optimal scheduling policy. Our obtained numerical results show that the proposed AA-DRL approach achieves improved PAoI performance compared to a recent benchmark~\cite{bench} while maintains lower complexity against the conventional iterative optimization method.

The remainder of the paper is organized as follows. Related work is presented in Section~\ref{Related}. Section~\ref{system-model} describes the system model. In Section~\ref{DRL}, we introduce our proposed DRL algorithm and the NN structure. Performance evaluations and conclusion are presented in Sections~\ref{results} and Section~\ref{sec:conclusions}, respectively.

\section{Related Work}
\label{Related}
Several research works have been conducted with the goal of minimizing AoI in wireless networks. The authors in ~\cite{q_management} studied the average AoI under different queue management schemes which concluded that packet replacement can promote reduced AoI compared to the conventional First Come First Served (FCFS) approach. The authors in \cite{schedule_2, schedule_4, schedule_5} introduced centralized scheduling methods to minimize the average AoI. However, the centralized scheduling approach is inefficient in D2D-enabled IIoT networks as it would incur high overhead and extended delay. Efficient scheduling in D2D networks was tackled by different works that focus on the analysis and optimization of resource allocation~\mbox{\cite{D2D_resource}}, traffic density~\mbox{\cite{D2D_traffic}}, or user fairness~\mbox{\cite{D2D_fairness}}, while less attention was paid to the minimization of AoI. Moreover, other works apply stochastic geometry to model the spatial relationship of D2D devices, and adopting AoI-aware decentralized scheduling with the assumption of having full CSI~\cite{stoch_1}. However, it is shown that it is very challenging to obtain a global CSI in D2D networks~\cite{edge2}.  A backlog-aware protocol was presented in~\cite{Q-aware} to minimize the average AoI subject to a delay constraint. The work in \cite{bench} presented a locally-adaptive slotted-ALOHA protocol where a link-wise access probability is dynamically selected to minimize the AoI considering a unit-size buffers and Last Come First Served (LCFS) queuing discipline. Both \cite{Q-aware} and \cite{bench} require the exchange of queue-status information between neighboring nodes to find the optimal access probability, which implies significant overhead and complexity in dense IIoT networks. Different from these works, our proposed AA-DRL has the potential to optimize the AoI performance of IIoT networks while maintaining low complexity with no explicit need for CSI.  

   \begin{figure*}[t!] 
		\centering
		\includegraphics[width= 0.7\linewidth]{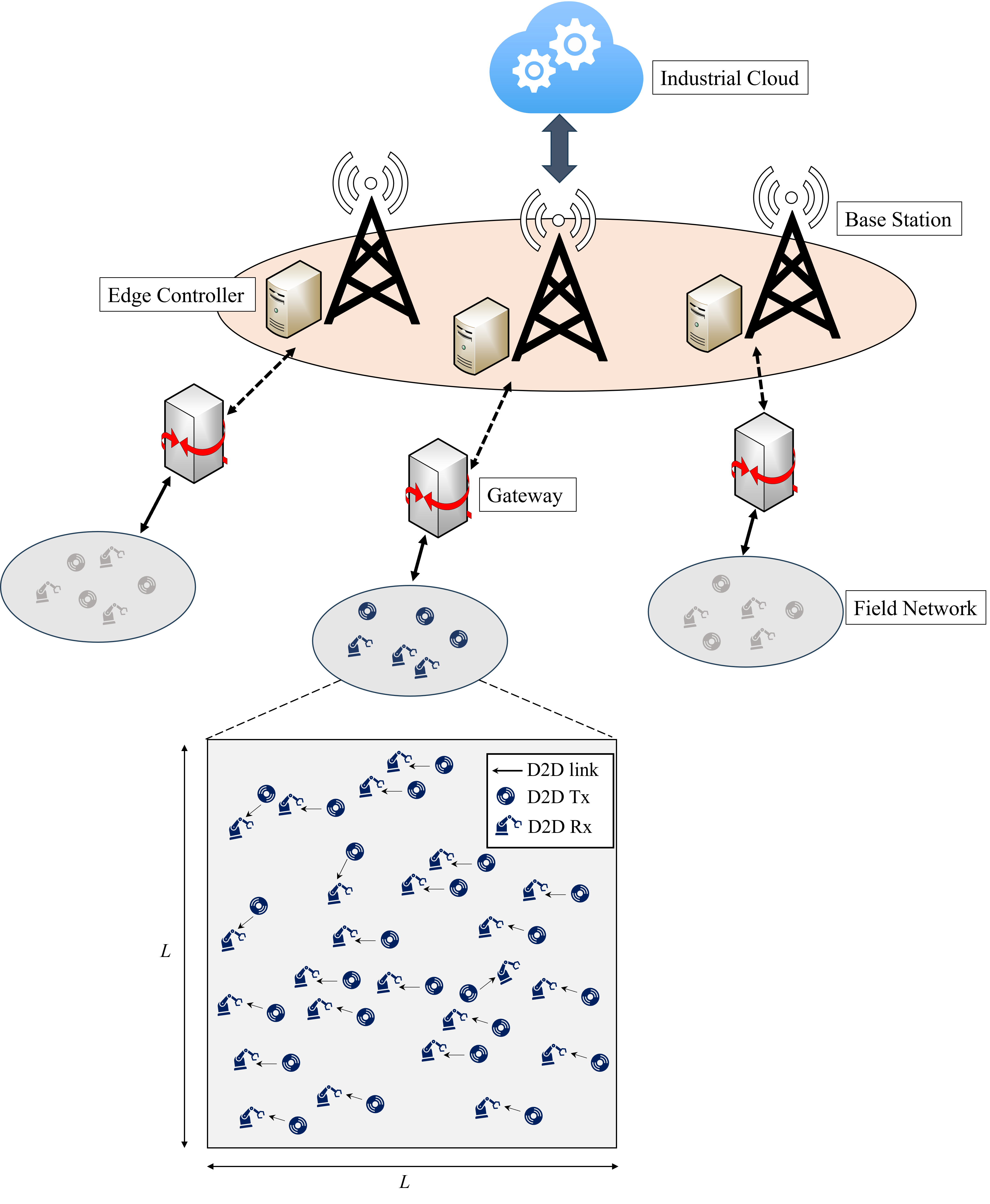}
		\caption{The hierarchy of the considered IIoT network.  \label{model}}
	\end{figure*}
\section{System Model}
\label{system-model}

We consider an IIoT network with a three-layer hierarchy as depicted in Fig.~\ref{model}. The first layer is the field network that consists of $N$ D2D pairs transmitting status updates via a shared time-slotted channel, where each transmission attempt fits within the duration of one time slot. The second layer includes edge computing nodes with storage, computing, communication, and other resources that are used for computing-intensive tasks. The third layer is the industrial cloud where historical data about the field and edge nodes is stored in the cloud for long-term data analysis. The considered model is common in most remote data acquisition and distributed control applications, and is shown to have promising advantages in supporting efficient resource management~\cite{netwModel}. In the field network layer, each transmitter generates status update following a Poisson process with average rate $\lambda$ packets/slot. Such arrival model captures the scenario where such traffic can be triggered by the occurrence of some random incident~\cite{arrival}. The D2D transmitters have a single-occupancy backlog, i.e., the output buffer can accommodate only one packet. Moreover, we consider a preemptive queuing policy where an arriving packet can preempt the one currently in the service (if there is such). We consider a stationary randomized scheduling policy \cite{R_policy}, in which the D2D transmitters are activated to transmit in  each time slot with a given slot-access probability that is subject to optimization. Let $\pi = \{p_1, p_2, ..., p_N\}$ denotes the the randomized scheduling policy and $\Pi$ is the class that represents all the possible policies where $\pi\in \Pi$. Particularly,  for a given scheduling policy $\pi$, the D2D transmitter $n_i$, $i \in \{1, 2, ..., N\}$, is activated with a probability $p_i$ across all the time slots. 
\subsection{Successful Update Probability}
The collisions among simultaneously active D2D pairs are not necessarily destructive, due to the capture effect~\cite{capture}. In that sense, a packet is decoded successfully at the end of a time slot when the Signal-to-Interference plus-Noise ratio (SINR) at the corresponding receiver exceeds the capture ratio $\beta$. The SINR at the AP in a time slot $t$ given a set $\mathcal{V}$ of interfering links can be written as~{\cite{stoch_1}}
\begin{equation}\label{SINR}
{\text{SINR}_i(t)= \frac{P_{t}|h_{i,i}|^{2}d_{i,i}^{-\alpha}}{\sigma^2+\sum_{j\in \mathcal{V}\backslash \{i\}}P_t|h_{j,i}|^{2}d_{j,i}^{-\alpha}}},  
\end{equation}
where $P_{t}$ is the transmission power (assumed fixed for all D2D pairs), $h_{j,i}$ is the random variable that represents the Rayleigh fading of the channel between transmitter of link $j$ and the receiver of link $i$ with $h_{j,i}\sim\mathrm{exp} (1)$, $d_{j,i}$ is the distance between  between transmitter of link $j$ and the receiver of link $i$,  $\sigma^2$ denotes the power of AWGN, and  $\alpha$ is the path loss exponent. According to the considered channel model, for an arbitrary node $n_i$, the conditional successful decoding probability in timeslot $t$ given a certain scheduling policy $\pi$ can be obtained as 
\begin{equation}\label{sucess_1}
\begin{split}
\phi_i(t|\mathcal{\pi})&=\mathrm{Pr(SINR}_{i}>\beta\,|\pi)\\
&=\mathrm{Pr}\left(\frac{P_{t}|h_{i,i}|^{2}d_{i}^{-\alpha}}{\sigma^2+\sum_{j\in \mathcal{V}\backslash \{i\} }P_t|h_{j,i}|^{2}d_{j}^{-\alpha}a_j(t)}>\beta|\pi\right)\\
&= \mathrm{exp}\left(-\frac{\beta\sigma^2}{P_{t}d_i^{-\alpha}}\right)\prod_{j\in \mathcal{V}\backslash \{i\}}\left(\frac{1}{1+a_j(t)\frac{\beta d_{j}^{-\alpha}}{d_{i}^{-\alpha}}}\right).
\end{split}
\end{equation}
Using \eqref{sucess_1} and considering the adopted randomized scheduling policy, the average successful probability $ \overline{\phi_i(t)}$ is given as 
\begin{equation}\label{sucess_2}
\overline{\phi_i(t)}=p_i\mathrm{exp}\left(-\frac{\beta\sigma^2}{P_{t}d_i^{-\alpha}}\right)\prod_{j\in \mathcal{V}\backslash \{i\}}\left(1-\frac{p_j}{1+\frac{d_{i}^{-\alpha}}{\beta d_{j}^{-\alpha}}}\right).
\end{equation}
\subsection{Analysis of the Peak AoI}
   \begin{figure}[t!] 
		\centering
		\includegraphics[width= 0.95\linewidth]{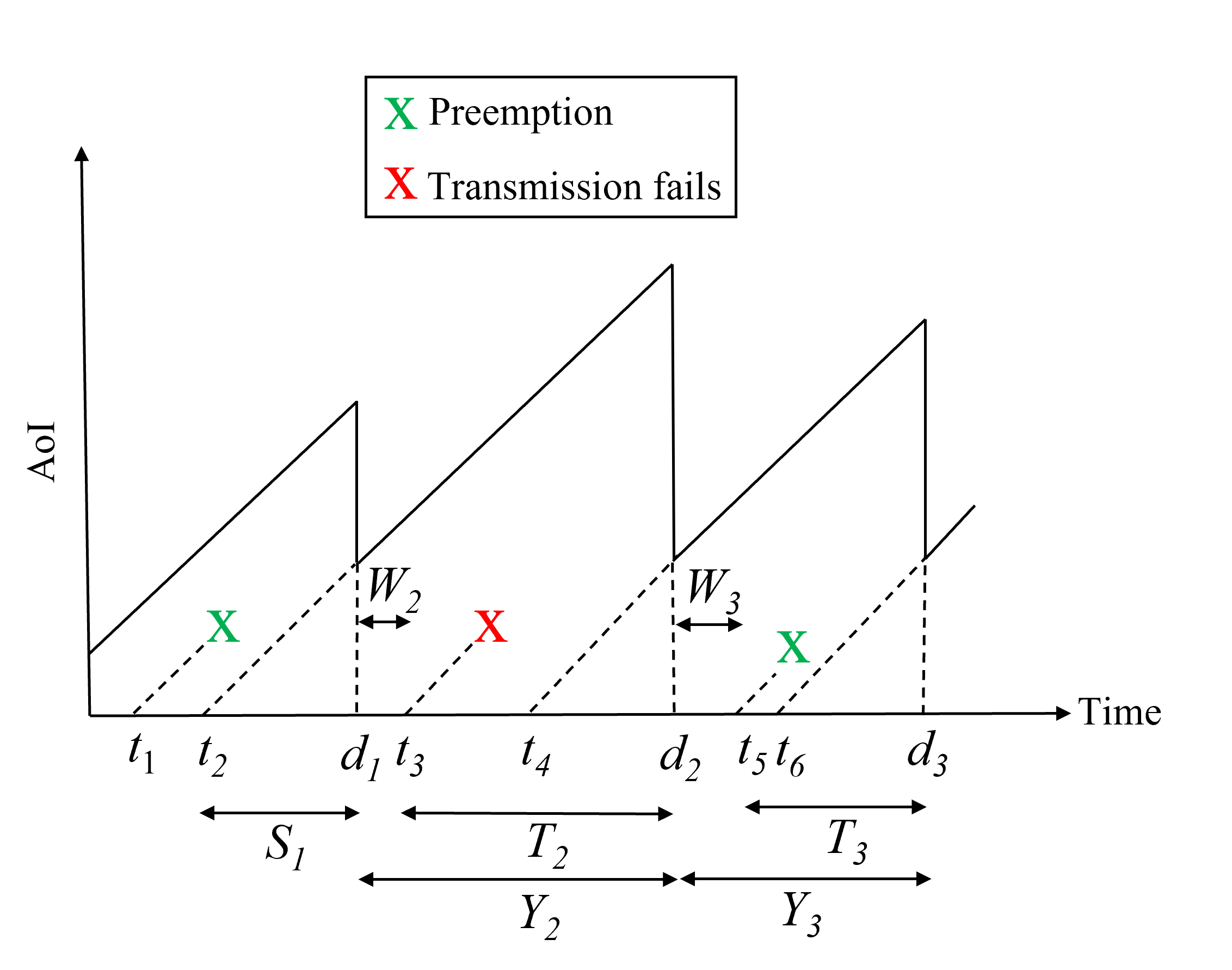}
    \vspace{-4mm}
		\caption{Evaluation of the AoI of an arbitrary node $n_i$ with $D = 6$.  \label{AoIEV1}}
	\end{figure}
The evaluation of the PAoI is statistically identical for all the D2D pairs, hence, in the following we focus on the derivation of the PAoI of an arbitrary D2D pair. Fig.~\ref{AoIEV1} shows an example evolution of the AoI of the considered preemptive queuing scheme.
Let $t_j$, $j=1, 2, 3, ....,$ the generation time of the $j$th update. We denote $X_j$ as the random variable that represents the interarrival time between consecutive updates, $X_j=t_{j+1}-t_j$, which follows an exponentially distribution with mean $1/\lambda$. Note that an update may not be received correctly by the D2D receiver due to transmission failures or preemption. Hereafter, we use a different index $i\leq j$ to refer to the successfully received updates. Let $d_i$ refers to the departure time of $i$th update that is successfully received by the corresponding D2D receiver, and $S_i$ is its corresponding service time. We denote $g_i$ as the generation time of the first generated update after $d_{i-1}$, and is given as
\begin{equation} 
g_i\triangleq \mathrm{min}\{t_j\mid t_j>d_{i-1}\}.
\end{equation}
Therefore, we can see that the indices $i\,\textrm{and}\,j$ in general do not refer to the same update.
For instance, in Fig.~\ref{AoIEV1}, the generated update at $t_3$ is not received, and the successfully received update at $d_2$ is the one generated at $t_4$.
We define $W_i=g_i-d_{i-1}$ as the interval between the reception of the $(i-1)$th update until the generation of the next update. We define the interval $T_i=d_i-g_i$, which represents the interval from $g_i$ until the next update received successfully. Note that $T_i$ spans the generation instants of failed updates. We also define the interdeparture time between two consecutive successfully received status updates $Y_i=d_i-d_{i-1}$. From the definition of $W_i$ and $T_i$, we have $Y_i=T_i+W_i$. Therefore, the PAoI , denoted as $A_{P_i}$ (the value of AoI just before receiving the update at $d_i$), can be given as
\begin{equation} 
A_{P_i}=Y_i+S_{i-1},
\end{equation}
where $S_{i-1}$ represents the service time of the update received before $d_i$. For instance, in Fig.~\ref{AoIEV1}, the PAoI at $d_2$ is equal to $Y_2+S_{1}$, where $S_{1}$ is the service time of the update received before $d_2$, which is $d_1$. The average PAoI ($\E[A_{P_i}]$) is given by 
\begin{equation} \label{PAoI} 
\E[A_{P_i}]=\E[Y]+\E[S]=\E[T]+\E[W]+\E[S],
\end{equation}
where $\E[Y]=\E[T]+\E[W]$. We denote $\gamma$ as the probability that an update is preempted. The value of $\gamma$ is given as a function of the arrival intensity and the service time as follows
\begin{equation} \label{preemp}
\gamma=1-e^{(-\mu\lambda)},
\end{equation}
where $\mu$ is the duration of one time slot, which represents the deterministic service time $S$ of an update.

For the considered queuing system, when a packet departs, it leaves the system empty, hence, $W$ will follow the same distribution as the interarrival time, i.e., $\E[W]=1/\lambda$. Moreover,  we have $\E[S]=\mu$. Therefore, the term $\E[T]$ can be evaluated using the following recursive method \cite{recursive} as
\begin{equation} \label{T}
\begin{split}
   \E[T]=&\underbrace{(1-\gamma)(1-\alpha)\E[S]}_{R_1}\\
   &+\underbrace{(1-\gamma)\alpha\left(\E[S]+\E[W]+\E[\hat{T}]\right)}_{R_2}\\
   &+\underbrace{\gamma(\E[X \mid X<S]+\E[\hat{T}])}_{R_3}, 
\end{split}
\end{equation}
where $\alpha = 1- \overline{\phi_i(t)}$ represents the failed transmission probability at time slot $t$. 
The first term $R_1$ in \eqref{T} denotes the case when the first update (generated at $g_i$) is not preempted ($1-\gamma$) by other updates and is received successfully. The term $R_2$ refers to the case when the first update is not preempted, but the transmission fails. In this case, the system spends the service time $S$ for the first update, then waits for the period $W$ until the next update is generated. 
The evaluation of $\hat{T}$ is the same as $T$, hence $\E[\hat{T}]=\E[T]$.
The term $R_3$ represents the case that the first generated update is preempted by a new update. In that case, the effective generation interval, i.e., the generation interval given that a packet, is preempted can be given as
\begin{equation} \label{AA}
\E[X\mid X<S]=\frac{\int_{0}^{\mu} s\lambda e^{-s\lambda }\,ds}{1-e^{(-\mu\lambda)}}=\frac{1}{\lambda}+\mu\left(1-\frac{1}{\gamma}\right).
\end{equation}
Using \eqref{AA} in \eqref{T} and substituting $\E[S]$ and $\E[W]$, $\E[T]$ can be obtained as
\begin{equation} \label{K2}
\E[T]=\frac{\gamma+\alpha-\gamma \alpha}{\lambda(1-\beta)(1-\alpha)}.
\end{equation}
Then, $\E[Y]$ becomes
\begin{equation} \label{Y}
\begin{split}
  \E[Y]&=\E[T]+\E[W]\\
  &=\frac{\gamma+\alpha-\gamma \alpha}{\lambda(1-\gamma)(1-\alpha)}+\frac{1}{\lambda}=\frac{1}{\lambda(1-\gamma)(1-\alpha)}.  
\end{split}
\end{equation}
Based on \eqref{PAoI} and \eqref{Y}, we obtain the average PAoI for the PR scheme $\E[A_{P_i}]$ as
\begin{equation} \label{without_ret}
\E[A_{P_i}]=\frac{1}{\lambda(1-\gamma)(1-\alpha)}+\mu.
\end{equation} 
For a given $\lambda$ and $\mu$, $\E[A_{P_i}]$ is mainly influenced by $\overline{\phi_i(t)}$, which depends on $p_i$. Our goal in this work is to find the optimal scheduling policy $\pi$ that minimizes the average PAoI $\E[A_{P_i}]$. Therefore, our optimization problem can be formulated as follows
\begin{equation}\label{opti_p1}
\begin{split}
&\underset{\pi \in \Pi}{\mathrm{min}}\,\,\, \E[A_{P_i}] \\
& \mathrm{s.t.}\,\,\, 0<p_i\leq 1,\,\, \forall i\in\{1,2, ...., N\}.\\
\end{split}
\end{equation}
Deriving a closed for expression for the solution of \eqref{opti_p1} a hard problem due to the spatio-temporal correlation of link states. In the following section, we introduce a DRL algorithm to solve optimization problem in \eqref{opti_p1}.

\section{The GLI-Based Deep Learning Approach}
\label{DRL}
\begin{figure}[t!] 
		\centering
		\includegraphics[width= 1\linewidth]{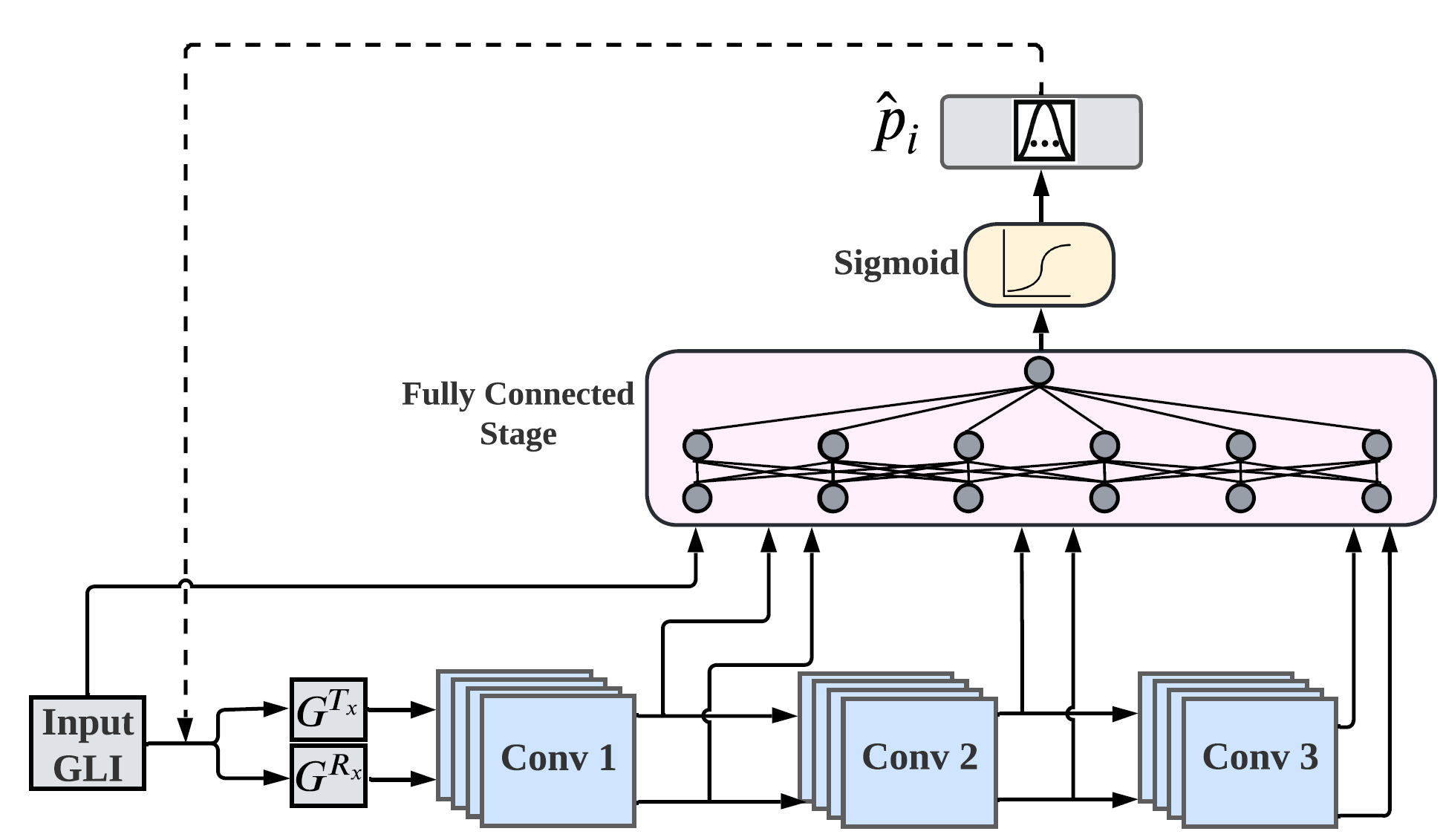}
    \vspace{-6mm}
		\caption{The structure of considered NN.  \label{NN}}
    \vspace{-4mm}
	\end{figure}

In this section, we present the design structure of the GLI-based NN to approximate the solution of \eqref{opti_p1} by mapping the GLI to the scheduling policy. The conventional method to solve the optimization problem in \eqref{opti_p1} requires a full information about the CSI, which incurs $O(N^2)$ computational complexity for a network of $N$ D2D pairs. The CSI could be mapped using GLI, which is considered as a function of CSI that captures the main feature of the wireless channels (the path loss and shadowing of a wireless link are mostly functions of distance and location). In that sense, we use the GLI as input to the NN to acquire the optimal scheduling policy $\pi^*$. The structure of NN is depicted in Fig.~\ref{NN} and is illustrated in more details as follows.
\subsection{D2D Density Grid}
\begin{figure}[t!] 
		\centering
		\includegraphics[width= 1\linewidth]{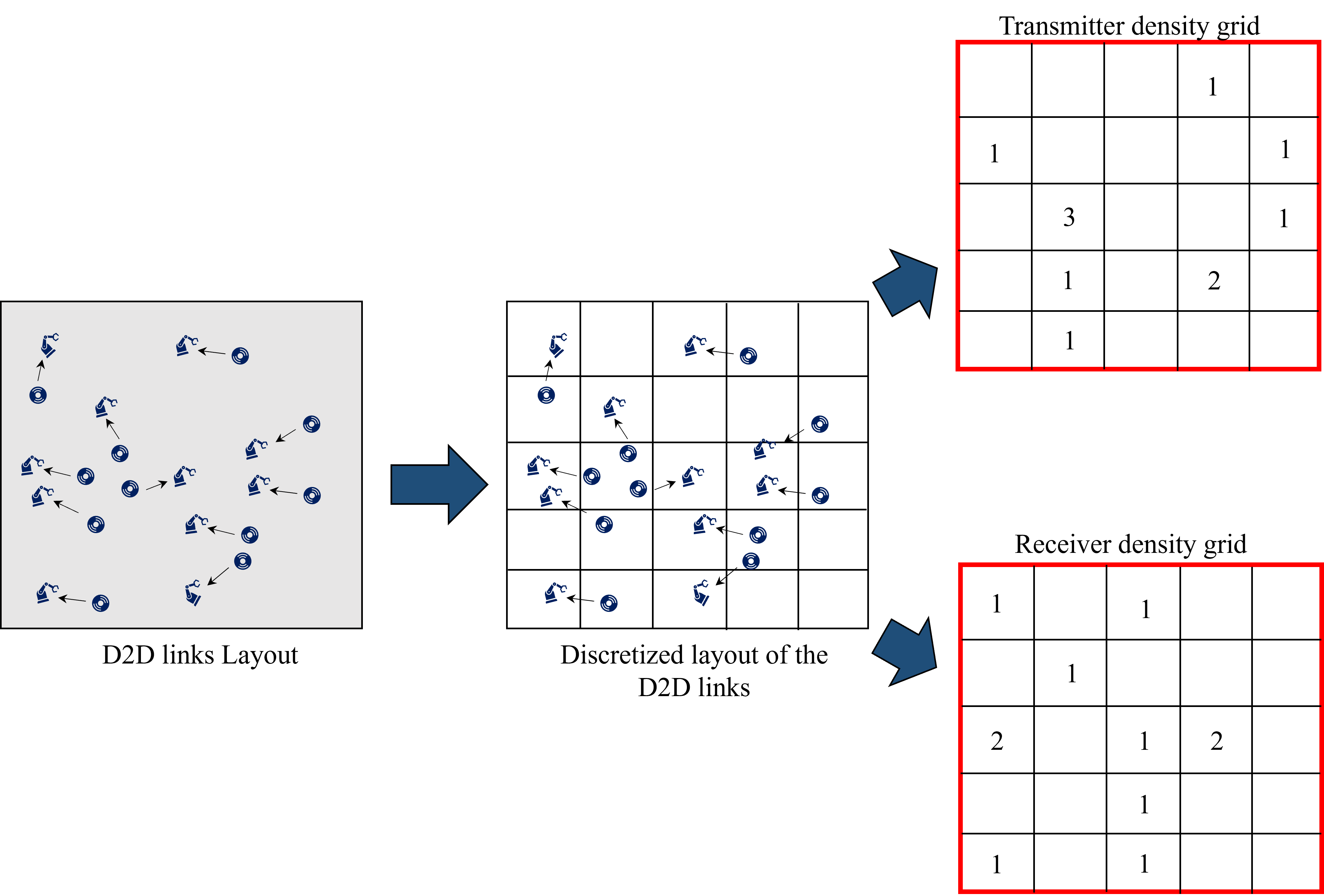}
    \vspace{-4mm}
		\caption{Construction of the transmitter and receiver density grids.  \label{grid}}
   \vspace{-4.5mm}
	\end{figure}

First, we construct grid matrices to quantize the continuous form of the locations of the transmitters and receivers as shown in Fig.~\ref{grid}. We assume that the considered network is distributed in a square-shaped area with a side length of $L$. The whole layout is then partitioned into square cells, where the GLI information is represented as the tuple $\{(x_i^{tx}, y_i^{tx}), (x_i^{rx}, y_i^{rx})\}_{i=1}^N$, where $(x_i, y_i)$ is the index of the cell and the coordinate values ranges from 0 to $L$. We consider two sub-grids, $G^{Tx}$ and  $G^{Rx}$ that represent the activation state of the transmitters and receivers, respectively. For a grid size of $R\times R$, the transmitter matrix $G_i^{Tx}$ of link $i$ is defined as
\begin{equation}
G_i^{Tx}=\begin{cases}
1 & \text{if} (x,y)=\lceil(x_i^{tx}, y_i^{tx})*R/L\rceil\\
0 & \text{otherwise.}
\end{cases}
\end{equation}
Based on the activation probability $p_i$, we have
\begin{equation}\label{GTX}
G^{Tx}=\sum_{j=i}^N p_j G_j^{Tx}.
\end{equation}
Note that the same applies to  $G^{Rx}$. In this case, the matrices $G^{Tx}$ and $G^{Rx}$ represent GLI information to solve the problem in \eqref{opti_p1} where probability $p_i$ reflects the interference to other links, hence they could be regarded as feature matrices to the convolution layer presented in the next subsection. 
\subsection{The 3-Layer Convolution Stage}
 The two  matrices $G^{Tx}$ and $G^{Rx}$ are processed via three connected convolution layers and the output is a set of extracted feature after each layer. Each entry in the resulting matrix comes from a unique convolution positioned at the corresponding index of the input matrix using the convolution filter. When the index represents the receiver's location, then the convolution essentially extracts features from all the transmitters in proximity to this receiver, based on the size of the convolution filter. Both $G^{Tx}$ and $G^{Rx}$ undergo the convolution phase concurrently, producing three matrices each. Then, every link on the D2D plane gleans a total of six features from these matrices, based on the index of its individual receiver or transmitter.
\subsection{The Fully Connected Stage}
The second stage is fully connected stage that comprises two hidden layers. We consider the Rectified Linear Unit (ReLU) as the activation function of each neuron where a sigmoid non-linearity is used at the output node to produce the activation probability $p_i$. For the considered D2D configuration with $N$ links, the feature vectors for each link are processed through the fully connected layer. This results in a collection of activation probability vectors $\pi$. In order to consider the Tx-Rx features in the learning process, we include the $p_i$ from the previous iteration and the distance between the Tx and Rx of the link as two features into the fully connected stage. Particularly, the output probabilities $p_i$ are used as inputs to get new GLI using \eqref{GTX}. This in turn helps to improve the convergence of the NN and enhance the training process in general.  
\subsection{The NN Training Process}
We train the NN using a randomly generated set of D2D layouts to minimize the AoI via gradient descent on the convolutional filter weights and the NN weight parameters. Specifically, the locations of $N$ transmitters are first generated uniformly within the region $L \times L$, and then the locations of the corresponding receivers are generated following a uniform distribution within a pairwise distances of $\{d_{min}, d_{max}\}$. Although the training stage would require the channel gains, it will no longer be required after the network is well-trained and only $N$ GLI is required to obtain the optimal schedule. This way, for each iteration, the NN tends to improve the scheduling policy of the previous iteration.  
\begin{table}[t!]
		\centering
		\caption{Evaluation parameters}
		\label{t1}
		\begin{tabular}{ll}
			\toprule
			Parameter & Value \\
				\midrule
			Deployment area & $600$ meters $\times$ $600$ meters\\
            Hidden layer & $30$ neurons\\
            GLI grid length $R$ & $150$\\
            Size of convolution filter & $10\times 10$\\
			Path loss exponent ($\alpha$) & \num {3} \\
			SINR threshold ($\beta$) & \num {0} dB \\
			Noise power ($\sigma^2$) & \num {-90} dBm\\
			Transmission power & \num {100} mW\\
			\bottomrule
		\end{tabular}	
	\end{table}
\section{Performance Evaluations}
\label{results}

In this section, we present the setup of the training and test processes of the considered NN, and evaluate the performance of the proposed AA-DRL via numerical results with the relevant parameters listed in Table~\ref{t1}.  

\subsection{Network Setup and Training Process}

We consider a D2D-assisted IIoT network where $N$ D2D pairs are distributed in $600$ meters $\times$ $600$ meters region. The  D2D transmitters are uniformly positioned within the deployment area while the locations of the corresponding receivers are generated following a uniform distribution within the pairwise distances $\{d_{min}=2$ meters and $d_{max}=80$~meters$\}$. We generate 10000 sets of such layout to train the NN and 5000 sets for testing. The NN consists of 3 convolution filters with each filter of size $20\times$ $20$. Each hidden layer in the fully connected stage comprises 30 neurons utilizing ReLU and sigmoid functions.  
   \begin{figure}[t!] 
		\centering
		\includegraphics[width= 0.95\linewidth]{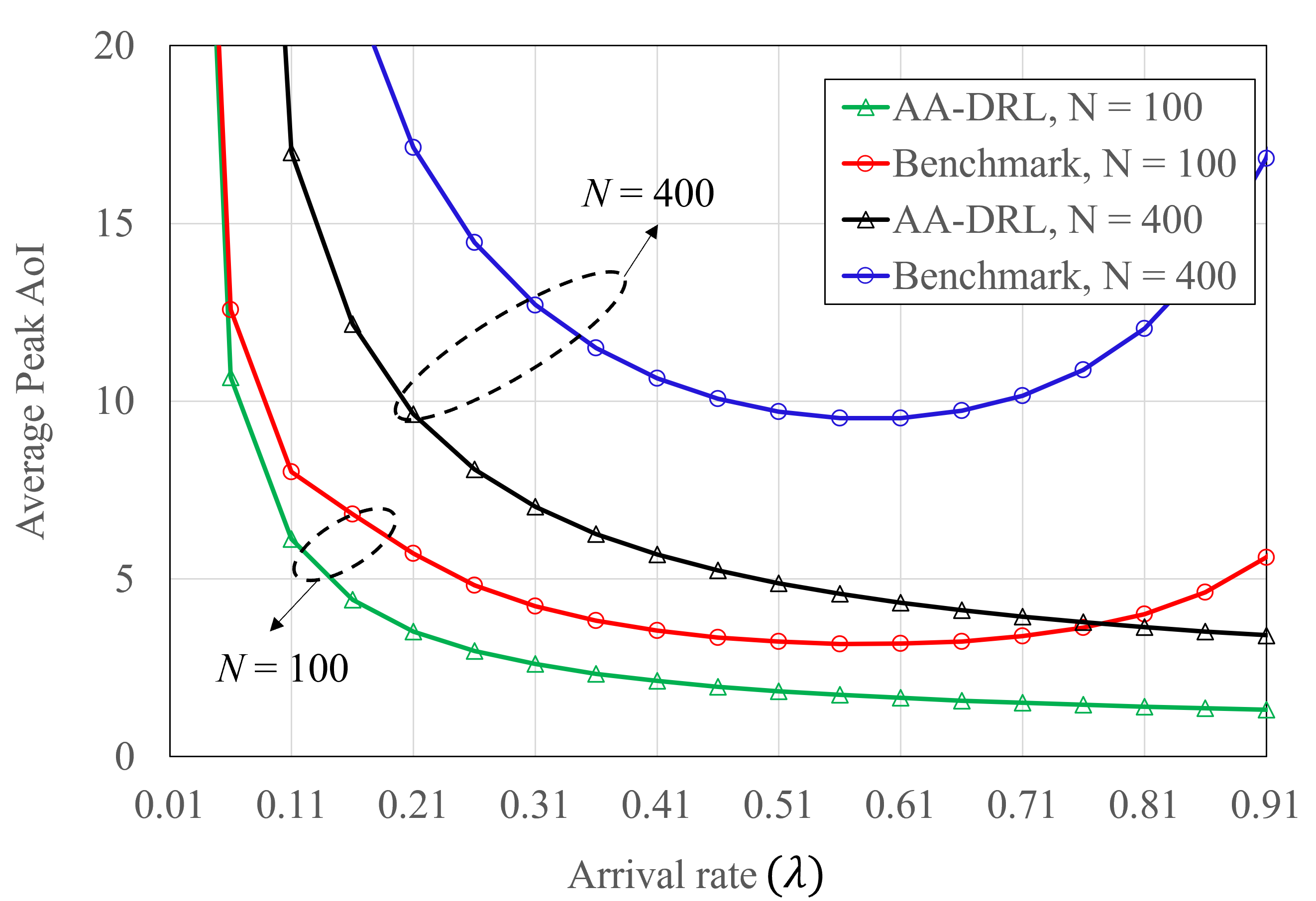}
		\caption{Comparison of the average PAoI under different values of $\lambda$ with $N = 100$ and $N = 400$.  \label{lambda}}
	\end{figure}
 \begin{figure}[t!] 
		\centering
		\includegraphics[width= 0.9\linewidth]{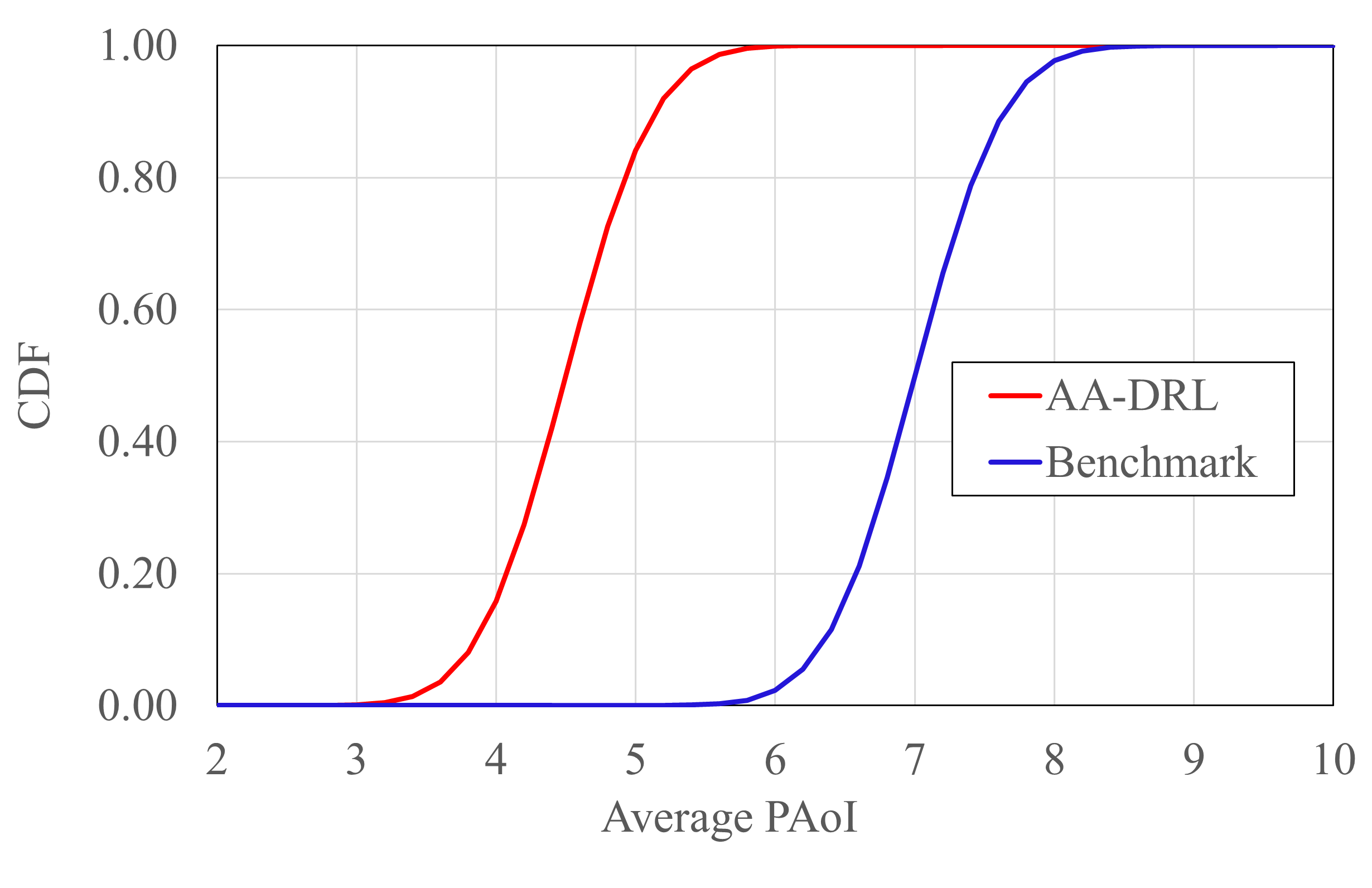}
    \vspace{-4mm}
		\caption{CDF of AA-DRL and benchmark with $N=100$ and $\lambda=0.2$ of 1000 different network layouts.  \label{CDF}}
	\end{figure}
\subsection{Numerical Results}
The following results show the performance of the proposed AA-DRL scheme in terms of the average PAoI and the computational complexity. Moreover, we use the work in~\cite{bench} as a benchmark to prove the effectiveness of our proposed approach. Fig.\ref{lambda} compares the average PAoI of AA-DRL and the benchmark in~\cite{bench} under varying $\lambda$. The figure shows superior performance of the proposed AA-DRL over the benchmark, especially when the network size increases from $N=100$ to $N = 400$. For instance, while AA-DRL achieves $55\%$ reduction in the AoI compared to the benchmark at $N=100$ and $\lambda=0.71$, this percentage increases to $75\%$ at $N=400$ and $\lambda=0.81$. The effectiveness of our proposed AA-DRL is attained through incorporating and mapping of the GLI in the scheduling policy, while the benchmark work is based only on local observations of the backlog status of the users, which would be inefficient in high interference regimes. 

To further emphasize the effectiveness of the proposed AA-DRL considering different layouts, we plot the Cumulative Distribution Function (CDF) of the average PAoI in Fig.~\ref{CDF} with $N=100$ and $\lambda=0.2$. Although different PAoI values are obtained under different layouts, we can observe that the CDF curves of both approaches have the same trend, hence the performance improvements of AA-DRL could be guaranteed under different network distributions (i.e., different spatial locations).
 \begin{figure}[t!] 
		\centering
		\includegraphics[width= 0.9\linewidth]{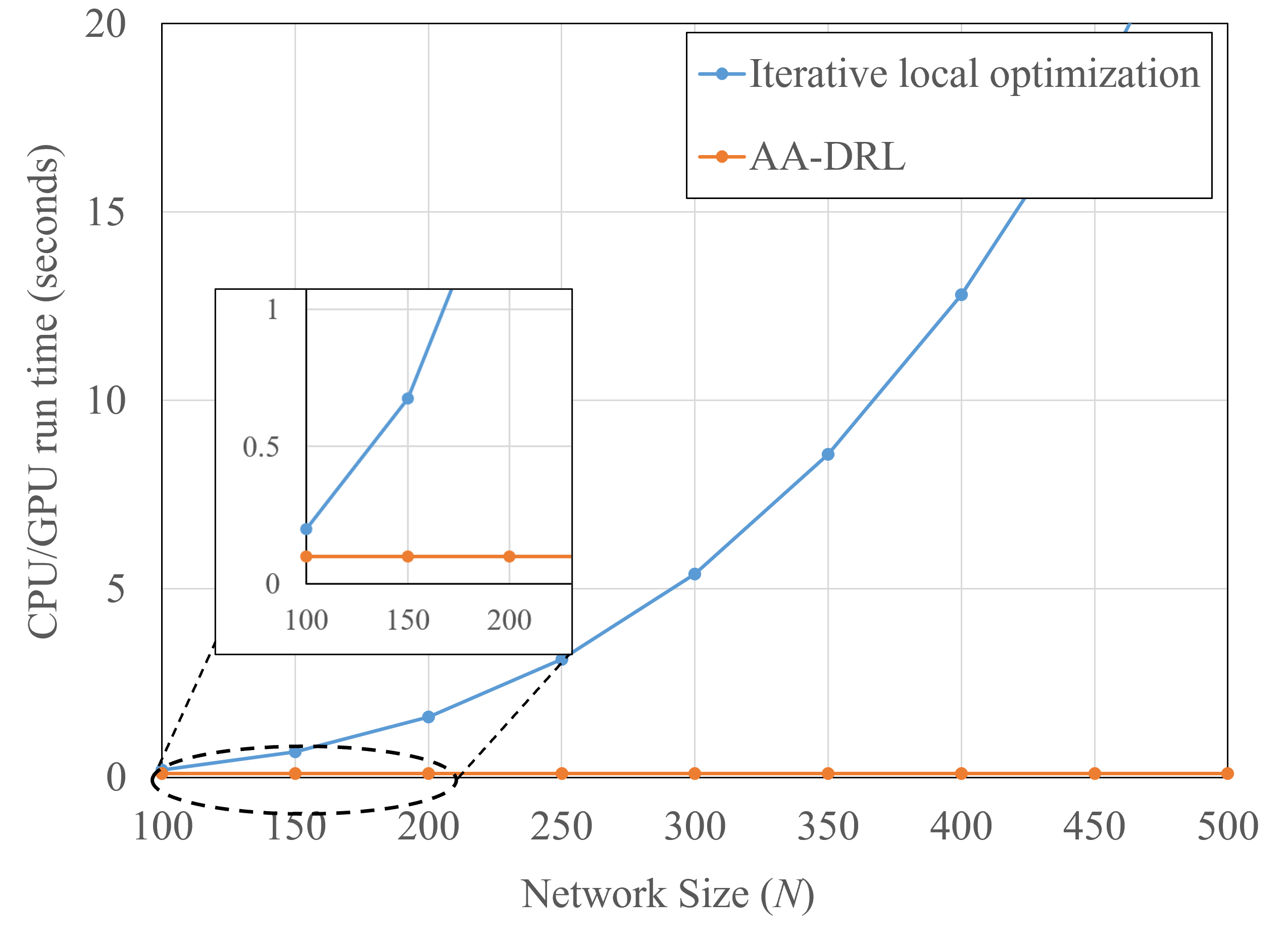}
    \vspace{-4mm}
		\caption{Log scale values of the computation time required to optimize a single layout under varying $N$.  \label{complex}}
	\end{figure}

One concern regarding the proposed AA-DRL would be its corresponding computational complexity. In the following, we roughly analyze the computational complexity of AA-DRL and compare it with an optimal iterative approach. The iterative approach uses a convex optimization
solver (e.g., CVX) to obtain a local optimal probability $p_i$. The access probability $p_i$ is updated in each iteration towards a minimum PAoI based on the obtained $p_i$ from the previous iteration. This way, $N$ one-variable problems are solved in a single iteration. For each iteration, the problem requires the collection of $N^2$ elements of the CSI matrix, leading to at least $O(N^2)$ computational complexity of this algorithm. For our proposed AA-DRL algorithm, the overall computation is $f\times[R^2\times(c_1^2+c_2^2+c_3^2)+N\times 10\times h_1 \times h_2]$, where $f$ is the number of feedback rounds, $c_i$ is the size of the convolution filter of stage $i$ and $h_1$ and $h_2$ are the number of neurons in each hidden layer. Therefore, for a given D2D layout,  the time complexity of AA-DRL scales as $O(N)$. Fig.~\ref{complex} shows the computation time in log scale of AA-DRL and the iterative method under different values of $N$. For the sake of reasonable comparison, we chose hardware configurations that are optimally compatible with each algorithm. For AA-DRL, we used Nvidia GPU GeForce GTX 1080Ti, while for the iterative algorithm we used Intel CPU Core i7-8700K @~3.70GHz. The design of the NN in the proposed AA-DRL is inherently amenable to parallel processing, gaining substantial advantages from the parallel computational capabilities of GPUs. Conversely, the iterative algorithm exhibits inherently sequential computation patterns, making it more suited to CPUs, which offer higher clock speeds. As it is demonstrated by Fig.~\ref{complex}, our proposed AA-DRL shows considerable computational advantages over the iterative approach in large-scale deployments while achieving optimal AoI performance. It is also worth mentioning that the considered IIoT architecture shown in Fig.~\ref{model} provides computational advantage where the complexity of AA-DRL would be further relaxed taking advantage of the computational capabilities of the edge node.

\section{Conclusion}
\label{sec:conclusions}

In this paper, we proposed a DRL-based approach to optimize AoI in D2D-assisted IIoT networks. We formulated a scheduling problem to minimize the average PAoI and developed a NN that maps the GLI to the optimal scheduling policy. The obtained results showed that our proposed approach achieves improved PAoI compared to a recent benchmark, all while exhibiting reduced computational complexity in contrast to the traditional iterative minimization algorithm. 

\section*{Acknowledgement}
This paper has received funding from the European Union’s Horizon 2020 research and innovation programme under grant agreement No. 883315.

	\bibliographystyle{IEEEtran}
\bibliography{mybib}
	
\end{document}